\documentclass{article}

\usepackage{graphicx}
\usepackage{graphpap,geometry}
\usepackage{amsmath}
\usepackage{amsfonts,amssymb}
\usepackage{cite}
\newcommand{\ket}[1]{| #1 \rangle}

\newcommand{\rb}[1]{\left( #1 \right)}

\title{{\bf  Exact isolated solutions for the two-photon 
Rabi Hamiltonian}}
\author{
C. Emary and R. F. Bishop\\
{\it Department of Physics},\\
{\it  University of Manchester Institute of
Science and Technology (UMIST)},\\
{\it P. O. Box 88, Manchester M60 1QD,
United Kingdom }}
\begin{document}
\parindent=0in
\maketitle
\begin{abstract}
\noindent
The two-photon Rabi Hamiltonian is a simple model describing the interaction
of light with matter, with the interaction being mediated by the exchange
of two photons. 
Although this model is exactly soluble in the rotating-wave 
approximation, we work with the full Hamiltonian, maintaining the
non-integrability of the model.  We demonstrate that, despite this 
non-integrability, there exist isolated, exact solutions for this 
model analogous to the so-called Juddian solutions found for
the single-photon Rabi Hamiltonian.  In so doing we use a Bogoliubov 
transformation of the field mode, as described by the present authors in
an earlier publication.

\end{abstract}

PACS number(s):  03.65.-w, 42.50.-p, 32.80.-t
\normalsize
\newpage

\section{Introduction}

The Rabi Hamiltonian (RH), introduced  by Rabi in 1937 \cite{ii:ra}, has long 
served as a popular and successful model of the interaction between
matter and electromagnetic radiation. The Hamiltonian 
provides a description of 
an atom approximated by a two-level system interacting via a dipole
interaction with a single mode of radiation \cite{al:eb}.  
Typically, this Hamiltonian
is studied within the rotating-wave approximation (RWA), which results in the
well-known Jaynes-Cummings model (JCM) \cite{ja:cu}.  The JCM is exactly 
integrable, whereas
the full RH is not.

The two-photon Rabi Hamiltonian (TPRH) is an obvious 
extension of the original RH, where the atomic transitions are 
induced by the absorption and emission of two photons rather than 
one.  Such non-linear optical processes have been of 
considerable interest [4], with applications including two-photon 
lasers and two-photon optical bistability [5].   The TPRH is not known 
to be integrable, whereas its RWA counterpart is, as has been 
demonstrated by Sukumar and Buck [6] and Compagno and Persico [7].
It should be noted from the outset that the TPRH, and its RWA variant, are 
phenomenological Hamiltonians, in that they neglect the effects
of intensity-dependent Stark shifts of the atomic levels \cite{pu:bu}.
Nevertheless they do provide useful prototypes of two-photon 
interactions \cite{ar:se}, and their
similarity with the RH and its RWA variant allows fruitful 
comparisons to be made \cite{fa:zh}.
The TPRH has been of considerable theoretical interest due to the connection 
of the two-photon interaction to the group $SU(1,1)$ and to the squeezed 
states \cite {cc:ge}.  Experimentally, the observation of two-photon Rabi 
oscillations in experiments with Rydberg atoms  \cite{ge:ba, ga:ba} 
has also contributed to the interest in this type of model.

Comparatively little attention has focused on the TPRH without the RWA.
A notable exception to this is the work of Ng {\it et al.} \cite{ng:lo},
who have 
used numerical diagonalisation 
in a truncated basis
to investigate the spectrum and 
simple dynamics of the system.  Their analysis indicates that the spectrum
of the full Hamiltonian is significantly different to the RWA spectrum, and 
that making the RWA also alters appreciably the dynamics.  These 
results fit in well with other work regarding the RWA, which 
suggest that the consequences of making this approximation may be greater than
usually thought \cite{fe:ra,fo:oc}.  Here we shall exclusively consider the 
full  Hamiltonian without the RWA.

In this communication we discuss a number of exact results for the TPRH
without the RWA.  After introducing the Hamiltonian and examining its 
connection to the group $SU(1,1)$, we consider the limit of the Hamiltonian in
which the atomic levels become degenerate.  In so doing we obtain a condition
on the range of atom-field couplings for which this model remains 
mathematically valid.    We then proceed to obtain a set of isolated, 
exact solutions for the Hamiltonian.  Their counterparts are well 
known for the single-photon RH, where they are referred 
to as Juddian solutions \cite{ju:dd, re:nu}.  Such solutions tell us 
a great deal 
about the structure 
and symmetries of this type of non-adiabatic model.  They may also serve as 
benchmarks for numerical techniques and as foundations 
for perturbative treatments.  Furthermore, the existence of isolated exact
solutions in 
non-integrable quantum models is also
of interest from the perspective of studying possible quantum chaos 
in such systems \cite{gr:ho, ho:ta}.
In determining 
these solutions for the TPRH we shall utilise an appropriate Bogoliubov 
transformation of the field mode, an approach outlined by the present 
authors in a previous publication \cite{ce:r1}, to be referred to 
as {\bf I} hereafter.

\section{The Hamiltonian}

The TPRH describes the interaction of a two-level atom with a single bosonic
field mode via a two-photon interaction.  The field is described by the 
annihilation and creation operators,
$b$ and $b^\dagger$ respectively, which obey the usual commutation
relation, $\left[b,b^{\dagger}\right] = 1$.
The two-level atom 
is described by the Pauli pseudo-spin
operators $\left\{ \sigma_k; k=x,y,z\right\}$, which satisfy 
the $SU(2)$ commutation relations,
$\left[\sigma_x, \sigma_y\right]=  2 i\sigma_z$, plus cyclic 
permutations.  We define
the raising and lowering operators as
$\sigma_+ \equiv \sigma_x + i \sigma_y$, 
$\sigma_- \equiv \sigma_x - i \sigma_y$.

In terms of these operators, the TPRH is given by
\begin{equation}
H_{2\gamma}=\frac{\omega_0}{2}\sigma_z + \omega b^\dagger b +
g\left(\left(b^{\dagger}\right)^2+b^2\right)\left(\sigma_+ +
\sigma_-\right),\label{TPRH}
\end{equation}
where $\omega_0$ is the atomic level splitting, $\omega$ is the
frequency of the boson mode and $g$ is the coupling strength of
the atom  to the field. Note that here we have the operators 
$\left( b ^\dagger\right)^2$ 
and $b^2$ inducing
atomic transitions, instead of $b^\dagger$ and $b$, as we would have in
the single-photon RH. It is convenient to rescale the Hamiltonian as 
$H_\mathrm{2\gamma} = \omega \tilde{H}_\mathrm{2\gamma}$, where
\begin{equation}
\tilde{H}_{\mathrm{2 \gamma}} = \tilde{\omega}\sigma_z + b^{\dagger}b + 
\lambda\left(\rb{b^{\dagger}}^2 + b^2 \right)\sigma_x,\label{TPRH2}
\end{equation}
and $\tilde{\omega} \equiv \frac{\omega_0}{2 \omega}$ and
$\lambda \equiv \frac{2g}{\omega}$.  
The TPRH is not known to be
integrable. 
Like the single-photon RH, the TPRH possesses a conserved
quantum number, which in the present case is 
$\pi_{2\gamma}$, namely an eigenvalue of the operator 
\begin{eqnarray}
\Pi_{2\gamma}&\equiv&\exp\left[\frac{i\pi}{2}\left(b^\dagger
                b+\sigma_z+1\right)\right] \\
             &=& - \sigma_z \exp\left[ \frac{i\pi}{2} b^\dagger b \right].
\label{TPpi}
\end{eqnarray}
It is simple to show that $\left[H_{2\gamma}, \Pi_{2\gamma}\right]=0$. 
The operator $\exp \rb{i \frac{\pi}{2} b^\dagger b}$ is the square 
root of the elementary parity operator 
$\exp \rb{i \pi b^\dagger b} =\Pi_{2\gamma}^2$,
and  has been denoted as the Fourier operator.  Its role in squeezing
has been described in detail elsewhere \cite{bi:v4}.

As the level-flip in the TPRH is induced by two photons, the
condition for the system to be on resonance is $\omega_0 = 2\omega$,
or alternatively that $\tilde{\omega} =1$.

\subsection{The TPRH and the $SU(1,1)$ Group}

The TPRH contains only quadratic combinations of the bosonic
creation and annihilation operators.  Consequently, we may write
the Hamiltonian in terms of the three operators $K_0$, $K_+$ and
$K_-$, which are defined as
\begin{equation}
K_+ \equiv \frac{1}{2}\left(b^\dagger\right)^2, ~~~~K_- \equiv
K_+^\dagger = \frac{1}{2}b^2, ~~~~K_0 \equiv \frac{1}{2}b^\dagger
b + \frac{1}{4}.
\end{equation}
These operators form a closed Lie algebra $SU\left(1,1\right)$,
defined by the commutator relations
\begin{equation}
\left[K_0,K_\pm\right]=\pm K_\pm,~~\left[K_-,K_+\right]=2K_0.
\end{equation}
The corresponding invariant Casimir operator $C$ is given by
\begin{equation}
C \equiv K_0^2 - \frac{1}{2}\rb{K_+K_- + K_-K_+},
\end{equation}
which commutes with all three generators.  For our purposes here,
we shall use a unitary irreducible representation of this algebra
known as the positive discrete series $\mathit{D}^+\rb{k}$
\cite{pe:re}.  In this representation the basis states $\left\{
\ket{k,m} \right\}$ diagonalise the operator $K_0$
\begin{equation}
K_0 \ket{k,m} = \rb{m+k} \ket{k,m},
\end{equation}
for $k>0$ and $m=0,1,2,\ldots$.  The action of the Casimir
operator in this representation is
\begin{equation}
C\ket{k,m} = k\rb{k-1}\ket{k,m}.
\end{equation}
The operators $K_+$ and $K_-$ are Hermitian conjugate to 
each other and act as
raising and lowering operators respectively within
$\mathit{D}^+\rb{k}$,
\begin{eqnarray}
K_+\ket{k,m} = \sqrt{\rb{m+1}\rb{m+2k}}\ket{k,m+1}, \nonumber \\
K_-\ket{k,m} = \sqrt{m\rb{m+2k-1}}\ket{k,m-1}.
\end{eqnarray}
For the single-mode bosonic realisation of $SU(1,1)$ that we require
here, the Bargmann index $k$ is equal to either $\frac{1}{4}$ or
$\frac{3}{4}$. In terms of the original Bose operators the states
$\ket{k,m}$ are given equivalently as
\begin{eqnarray}
|1/4,m\rangle \equiv \frac{1}{\left(2m\right)!}
  \left( b^\dag\right)^{2m} | 0 \rangle;
~~~~
|3/4,m\rangle \equiv \frac{1}{\left(2m+1\right)!}
  \left( b^\dag\right)^{2m+1} | 0 \rangle,~~~
m=0,1,2,\ldots.
\end{eqnarray}
So by switching to a $SU(1,1)$ representation we
explicitly acknowledge that we are splitting the Hilbert space of
the boson field into two independent subspaces.  Each subspace is
labeled by the Bargmann index $k=\frac{1}{4},\frac{3}{4}$ and only
contains either all even ($k=\frac{1}{4}$) or all odd
($k=\frac{3}{4}$) number states .
It is interesting to note in passing that the algebra $SU(1,1)$ may also be
used to describe a system of two bosonic modes, which interact in
such a way as to preserve the total particle number
\cite{bi:v3,cg:rg}

In terms of the $SU(1,1)$ generators, the TPRH can be written
\begin{equation}
H_{2\gamma}=\frac{\omega_0}{2}\sigma_z
                         + 2 \omega \left( K_0 - \frac{1}{4}\right)
                 + 2 g \left( K_+ + K_-\right)\left(\sigma_+ + \sigma_-\right),
\end{equation}
with the rescaled Hamiltonian being given by
\begin{equation}
\tilde{H}_{\mathrm{2 \gamma}} =\tilde{\omega} \sigma_z
                         + 2 \left( K_0 - \frac{1}{4}\right)
                 + 2 \lambda \left( K_+ + K_-\right)\sigma_x.
\end{equation}

\subsection{Squeezing and $SU(1,1)$}

The relationship between the group $SU(1,1)$ and squeezing has 
been described in detail elsewhere \cite{bi:v2} and the use of squeezed 
states in finding exact isolated solutions has been previously 
discussed in {\bf I}.  Here we simply note that the general squeezing 
operator $S$ can be written as
\begin{eqnarray}
S\rb{\sigma, \beta} = 
  \exp\rb{\sigma K_+} 
  \rb{1-|\sigma|^2}^{K_0} 
  \exp\rb{-\sigma^* K_-} 
  \exp\rb{\beta\rb{K_0-1/2}},
\end{eqnarray}
where $\sigma$ and $\beta$ are squeezing parameters, with $\beta$ real and
$\sigma$ a complex number with modulus $|\sigma| < 1$.  $S$ is a unitary 
operator, $S^\dagger S = 1$, and provides a representation of 
the group  $SU(1,1)$.  With it we may  make unitary transformations 
of the bosonic annihilation and creation operators, such that
\begin{eqnarray}
S\left( \sigma, \beta \right) b S^\dagger \left( \sigma, \beta
\right) &=& e^{-i\beta} \left( 1 - |\sigma|^2\right)^{-1/2} \left(
b - \sigma b^\dagger\right) \equiv c, \nonumber \\ S\left( \sigma,
\beta \right) b^\dagger S^\dagger \left( \sigma, \beta \right) &=&
e^{i\beta} \left( 1 - |\sigma|^2\right)^{-1/2} \left( b^\dagger -
\sigma^* b\right) \equiv c^\dagger. \label{SQcdefn}
\end{eqnarray}
The operators $c$ and $c^\dagger$ satisfy the commutation relation
$\left[c, c^\dagger \right]=1$, and are henceforth referred to as 
squeezed boson operators.

\section{Degenerate atomic levels}

For degenerate atomic levels, $\omega_0 = 0 = \tilde{\omega}$, 
the (rescaled) TPRH takes 
the form
\begin{equation}
\tilde{H}_{2\gamma}^{\rb{\omega_0=0}}= b^\dagger b + \lambda
\left(\left(b^{\dagger}\right)^2+b^2\right)\sigma_x.\label{TPRHom0a}
\end{equation}
Consequently eigenstates of $\tilde{H}_{2\gamma}^{\rb{\omega_0=0}}$ are
also eigenstates of $\sigma_x$, and we are led to consider the 
bosonic Hamiltonian,
\begin{equation}
\tilde{h}_{2\gamma}^{\rb{\omega_0=0}} =   b^\dagger b \pm
\lambda \left( \left(b^\dagger\right)^2 + b^2\right),
\label{TPMhom0}
\end{equation}
where the two signs correspond to the two eigenvalues of $\sigma_x$.
The Hamiltonian of Eq. (\ref{TPMhom0}) has the form of a 
squeezed harmonic oscillator.
In seeking its
eigen-solutions, it is convenient to use the squeezed bosons discussed 
above.

Inverting the relations (\ref{SQcdefn}), setting $\beta$ to zero and 
constraining $\sigma$ to be real, we obtain the following 
forms for the squeezed bosonic operators
\begin{equation}
c^\dagger_\pm \equiv \frac{b^\dagger \pm \sigma b}{\sqrt{1
-\sigma^2}},~~c_\pm \equiv \frac{b \pm \sigma
b^\dagger}{\sqrt{1-\sigma^2}},
\end{equation}
where the subscript on these operators corresponds to the sign in 
Eq. (\ref{TPMhom0}).
We now choose $\sigma$ to be given by
\begin{equation}
\sigma = \frac{1-\Omega}{2\lambda};~~~~\Omega \equiv
\sqrt{1-4\lambda^2}. \label{bigOm}
\end{equation}
Writing the Hamiltonian in terms of these squeezed operators with
this value of $\sigma$ we have
\begin{equation}
\tilde{h}_{2\gamma}^{\rb{\omega_0=0}} =  \Omega\rb{c_\pm^\dagger
c_\pm + \frac{1}{2}} - \frac{1}{2}.\label{TPOmno}
\end{equation}
The eigenstates of this Hamiltonian are clearly the number states
of the $c_\pm$-type bosons, which we denote $\ket{n,\mp\sigma}$, 
such that $c_\pm^\dagger c_\pm \ket{n,\mp\sigma}= n\ket{n,\mp\sigma}$.  In
our original unsqueezed representation, these states have the form
\begin{equation}
\ket{n;\mp\sigma} =
\frac{\rb{1-\sigma^2}^{1/4}}{\sqrt{n!}}\left[\frac{b^\dagger \pm
\sigma b }{\sqrt{1-\sigma^2}}\right]^n  
e^{\mp \frac{1}{2} \sigma {b^\dag}^2}\ket{0},\label{2pbefn}
\end{equation}
showing them to be the usual squeezed number states \cite{bi:v2}.  
Thus we see the eigenenergies of the Hamiltonian of 
Eq. (\ref{TPOmno}) to be
\begin{equation}
\tilde{E}^{(\omega_0=0)}_{n} =  \left\{ n +
\frac{1}{2}\right\}\Omega - \frac{1}{2},
\end{equation}
such that $\tilde{h}_{2\gamma}^{\rb{\omega_0=0}}\ket{n;\mp\sigma} 
= \tilde{E}^{(\omega_0=0)}_{n}\ket{n;\mp\sigma} $.

An important feature of the TPRH is revealed by considering
this $\omega_0 = 0$ case.  As we saw above, the eigenfunctions of
the bosonic part of the $\omega_0=0$ Hamiltonian are number states of
the squeezed bosons.  The squeezed vacuum, $\ket{0;\mp \sigma}$,
is proportional to $\exp\rb{\mp\frac{1}{2}\sigma
{b^\dagger}^2}\ket{0}$ and this state is only normalisable 
for $|\sigma|<1$, which corresponds to the conditions
\begin{equation}
|\lambda|< \frac{1}{2};~~~~~|\frac{4g}{\omega}| < 1.
\end{equation}
Above this value of $g/\omega$ the $\omega_0=0$ Hamiltonian does
not possess normalisable eigenfunctions and is thus unphysical.
As has been discussed by Ng {\it et al.} \cite{ng:lo}, and as is borne 
out by numerical diagonalisation \cite{ce:th}, this analysis 
still holds for the $\omega_0 \ne 0$
case as the remaining operator in the Hamiltonian,
$\tilde{\omega} \sigma_z$, is clearly a bounded operator
\cite{ma:ph}, and thus presents no problems.  Thus we see that the TPRH is
only well defined for values of $|\lambda|<1/2$.
This restriction on the coupling is not a severe restriction as, 
at higher couplings, effects not included in this model, such as the 
contributions of off-resonant, one-photon processes, will come into 
play, thereby compromising the physical relevance of the model.

\section{Isolated exact solutions}

We now demonstrate the existence of a class of
isolated, exact solutions for the TPRH, similar to the Juddian solutions
found for the one-photon RH. 
Following {\bf I} we seek solutions by first performing a 
Bogoliubov transformation of the field mode.  Bearing in mind 
the $\tilde{\omega}=0$ result, we shall make the transformation 
from the original bosons $b$ and
$b^\dagger$ to the squeezed bosons $c$ and $c^\dagger$,
\begin{equation}
b = \rb{1 - \sigma^2}^{-1/2} \rb{c + \sigma c^\dagger} ;~~
b^\dagger = \rb{1 - \sigma^2}^{-1/2} \rb{\sigma c + c^\dagger},
\end{equation}
where we have assumed that  $\beta=0$ and that $\sigma$ is real and
to be determined.  The justification of this
is provided by subsequent results.

With this change in bosonic representation, the rescaled TPRH becomes
\begin{eqnarray}
\tilde{H}_{2\gamma} = \tilde{\omega} \sigma_z &+&
\frac{1}{\kappa} \rb{ \sigma c^2 + \sigma {c^\dagger}^2  + \rb{1
+ \sigma^2}
c^\dagger c + \sigma^2  } \nonumber \\
&+& \frac{\lambda}{\kappa}\rb{ \rb{1 + \sigma^2} \rb{{c^\dagger}^2 +
c^2} + 4 \sigma c^\dagger c + 2 \sigma }\sigma_x,
\end{eqnarray}
where $\kappa \equiv \rb{1- \sigma^2}$. We now use an appropriate 
matrix representation
for the Pauli matrices, which  for this model is one 
in which $\sigma_x$ is diagonal.  We use
\begin{equation}
\sigma_x = \left[\begin{array}{lr}1 & 0\\0 & -1 \end{array} \right],~~
\sigma_y = \left[\begin{array}{lr}0 & i\\-i & 0 \end{array} \right],~~
\sigma_z=\left[\begin{array}{lr}0 & 1\\1 & 0 \end{array} \right]. 
\end{equation}
In terms of the two-component 
wavefunction, 
$\ket{\Psi} =  {\ket{\Psi_1} \choose \ket{\Psi_2}}$,
the time-independent Schr\"{o}dinger equation for the system,
$\tilde{H}_\mathrm{2\gamma}\ket{\Psi} = \tilde{E}\ket{\Psi}$, where 
$\tilde{E} \equiv E / \omega$ is the rescaled energy, then reads
\begin{eqnarray}
\tilde{\omega} \ket{\Psi_2} &+& \frac{1}{\kappa} \left\{\left[ \sigma +
\lambda \rb{1 + \sigma^2} \right] \rb{{c^\dagger}^2 + c^2} \right.\nonumber \\
&&\left. + \left[ 1 + \sigma^2 + 4\lambda \sigma \right]c^\dagger
c + \sigma^2 + 2 \lambda \sigma - \kappa \tilde{E} \right\}
\ket{\Psi_1} =
0, \label{2pJ2e1}\\
\tilde{\omega} \ket{\Psi_1} &+& \frac{1}{\kappa} \left\{\left[ \sigma -
\lambda \rb{1 + \sigma^2} \right] \rb{{c^\dagger}^2 + c^2} \right.
\nonumber \\
&&\left. + \left[ 1 + \sigma^2 - 4\lambda \sigma \right]c^\dagger
c + \sigma^2 - 2 \lambda \sigma - \kappa \tilde{E} \right\}
\ket{\Psi_2} = 0 \label{2pJ2e2}
\end{eqnarray}

It is immediately clear that if we set either $\left[ \sigma +
\lambda \rb{1 + \sigma^2} \right]$ or $\left[ \sigma - \lambda
\rb{1 + \sigma^2} \right]$ equal to zero, we make a determination
of $\sigma$ and reduce either Eq. (\ref{2pJ2e1})
or Eq. (\ref{2pJ2e2}) considerably.  Choosing the first of these
options, we have
\begin{equation}
\sigma + \lambda \rb{1 + \sigma^2}=0
\end{equation}
which gives
\begin{equation}
\sigma = \frac{-1 \pm \Omega} {2 \lambda}, \label{2psig1}
\end{equation}
where $\Omega$ is as defined previously in Eq. (\ref{bigOm}).  
In order that $|\sigma|<1$, we must henceforth choose
the positive sign in Eq. (\ref{2psig1}), so that $\sigma
\rightarrow 0$ as $ \lambda \rightarrow 0$.  Thus our squeezing
parameter is
\begin{equation}
\sigma = \frac{\Omega -1} {2 \lambda} \label{2psig2}.
\end{equation}

Note that had we pursued the other option and set 
$\left[ \sigma - \lambda\rb{1 + \sigma^2} \right] = 0$, we would 
have obtained the same determination of the squeezing parameter as for the 
$\omega_0 = 0$ case given by Eq. (\ref{bigOm}).  This second solution 
will be discussed later.
Proceeding with this choice of squeezing given by Eq. (\ref{2psig2}),
Eqs. (\ref{2pJ2e1}) and (\ref{2pJ2e2}) become
\begin{eqnarray}
\tilde{\omega} \ket{\Psi_2} &+& \left\{ \Omega c^\dagger c -
\left[\tilde{E} + \frac{1}{2} - \frac{\Omega}{2}\right] \right\}
\ket{\Psi_1} = 0, \\
 \tilde{\omega} \ket{\Psi_1} &+& \frac{1}{\Omega}\left\{
-\sqrt{1 - \Omega^2}\rb{{c^\dagger}^2 + c^2} + \rb{2-\Omega^2}
c^\dagger c \right.
\nonumber \\
&&\left. + \frac{1}{2}\rb{1- \Omega}\rb{2 + \Omega} - \Omega
\tilde{E}\right\} \ket{\Psi_2} = 0.
\end{eqnarray}

For $\ket{\Psi_1}$ and $\ket{\Psi_2}$ we now choose simple
Ans\"{a}tze in terms of the squeezed number states;
\begin{eqnarray}
\ket{\Psi_1} = \sum_{n=0}^{N} p_n \ket{n;\sigma}, \\
\ket{\Psi_2}  = \sum_{m=0}^{N-2} q_m
\ket{m;\sigma}\label{2pJSansatz},
\end{eqnarray}
which gives us the equations
\begin{eqnarray}
\sum_{m=0}^{N-2}\tilde{\omega} q_m \ket{m;\sigma}
&+&\sum_{n=0}^{N}  p_n 
\left\{
  n \Omega -
  \left[
    \tilde{E} +\frac{1}{2} - \frac{\Omega}{2}
  \right] 
\right\} 
\ket{n;\sigma} = 0, \label{2pzpowi}\\
\sum_{n=0}^{N} \tilde{\omega} p_n \ket{n;\sigma} 
&+& \sum_{m=0}^{N-2} q_m \frac{1}{\Omega} 
\left\{
  -\sqrt{1-\Omega^2} \left[\sqrt{\rb{m+1}\rb{m+2}}\ket{m+2;\sigma}
  \right. \right.\nonumber \\ &&\left. \left. + \sqrt{m
  \rb{m-1}}\ket{m-2;\sigma}\right]
  + \rb{2 - \Omega^2}m \ket{m;\sigma} \right. \nonumber \\
  &&\left. + \frac{1}{2} \left[\rb{1- \Omega}\rb{2 + \Omega} -
  2\Omega \tilde{E} \right] \ket{m;\sigma} 
\right\} = 0
\label{2pzpowii}
\end{eqnarray}
For the first equation to hold, we must have
\begin{equation}
p_N \left\{N \Omega - \left[\tilde{E} +\frac{1}{2} -
\frac{\Omega}{2}\right] \right\} = 0.
\end{equation}
As $p_N \ne 0$ by Ansatz, we must have $\Omega N - \rb{\tilde{E}
+ \frac{1}{2} - \frac{\Omega}{2}} = 0$.  This establishes the
so--called energy baselines as
\begin{equation}
\tilde{E} = - \frac{1}{2} + \rb{N + \frac{1}{2}}\Omega.
\end{equation}

Equating the remaining coefficients of the number states in
Eqs. (\ref{2pzpowi})  and (\ref{2pzpowii}) gives us 
the following set of equations
\begin{eqnarray}
\tilde{\omega} q_m &+& \rb{m - N} \Omega p_m =
0;~~~~~~~~~~~~~~~~~~~~~~~~~~~~~~~~~~~~~~0 \le m \le N-2,
\\
\tilde{\omega} p_n &-& \frac{\sqrt{1 -
\Omega^2}}{\Omega}\sqrt{n\rb{n-1}} q_{n-2} - \frac{\sqrt{1 -
\Omega^2}}{\Omega}
\sqrt{\rb{n+1}\rb{n+2}}q_{n+2}
 \nonumber \\
 &+& \frac{1}{\Omega} \left[ n +1 - \rb{2n+N+1} \Omega^2 \right] q_n = 0;
~~~~~~~~~~~~~~~~~~~~ 0\le n \le N.
\end{eqnarray}
where $q_n = 0$ for $n<0$ and  $n> N-2$.  From the second of these
equations we see that the Hamiltonian in this squeezed
representation only couples number states that are different by
multiples of two (for example it couples $\ket{n;\sigma}$ to
$\ket{n+2;\sigma}$ and $\ket{n+4;\sigma}$ but not to
$\ket{n+3;\sigma}$). Therefore our Ans\"{a}tze need only include either
all even number states or all odd number states. This is equivalent to 
restricting the solutions to a sector of the Hilbert space with a given 
Bargmann index, $k$.  We also see that
the minimum value of $N$ is 2. When $N$ is even, we obtain $N+1$
equations for the $N+1$ coefficients $p_n;~n=0,2,\ldots,
N,~~~q_m;~ m=0,2,\ldots N-2$. When $N$ is odd, we obtain $N$
equations for the $N$ coefficients $p_n; n=1,3,\ldots, N,~~q_m;
m=1,3,\ldots N-2$. Requiring that the determinants of these
equation sets are equal to zero gives us the compatibility
conditions which establish the location of the Juddian  points on
the energy baselines.  These conditions for the first three
values of $N$ are:
\begin{eqnarray}
2 - 6 \Omega^2 + \tilde{\omega}^2 &=& 0;~~~~N=2, \nonumber\\
6 -10 \Omega^2 + \tilde{\omega}^2 &=& 0;~~~~N=3, \nonumber\\
  8 \left(3 - 30 \Omega^2 + 35 \Omega^4 \right)
      + 2 \left( 7 - 17 \Omega^2\right)\tilde{\omega}^2
      + \tilde{\omega}^4 &=& 0
;~~~~N=4.
\end{eqnarray}
For $N$ even we have a polynomial of the
order $\frac{N}{2}$  in $\tilde{\omega}^2 $ and
$\Omega^2$, and hence $\lambda^2$.  For $N$ odd, the corresponding 
polynomial is of order ($\frac{N-1}{2}$). The equations are symmetric
under $\tilde{\omega} \rightarrow -\tilde{\omega}$ or $\lambda
\rightarrow - \lambda$, as expected.   The solutions of 
these polynomials locate the exact solution 
in  $\tilde{\omega}$--$\lambda$ space.

As in the one-photon case, there exists another degenerate
solution at every Juddian point.  This solution can be found by
placing $\left[ \sigma - \lambda \rb{1+\sigma^2}\right]=0$ rather
than $\left[ \sigma + \lambda \rb{1+\sigma^2}\right]=0$ into
Eqs. (\ref{2pJ2e1}) and (\ref{2pJ2e2}).  This simply interchanges the
equations and hence the roles of $\Psi_1$ and $\Psi_2$.

\section{Results and Discussion}

\begin{figure}[p]
\centerline{\includegraphics[height=4.5in]{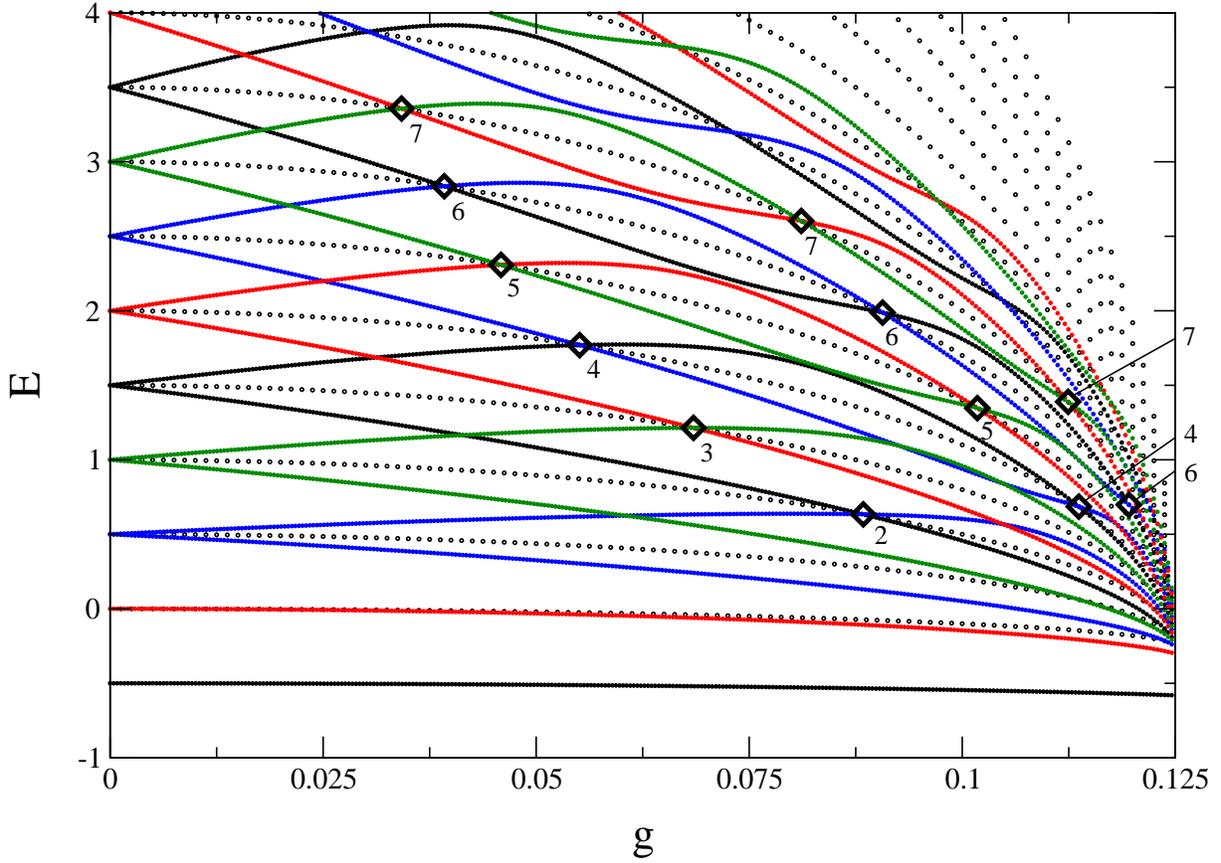}} \caption{
The first twelve Juddian points (diamonds) of the two-photon Rabi
Hamiltonian determined by the method outlined in the text, plotted
against the energy spectrum determined numerically (solid lines).  Also plotted are
the energy baselines (dotted lines).  Each point is labeled with its order 
$N$.  The Hamiltonian is resonant;
$2\omega = \omega_0 = 1$. \label{2pJuddfig1}}
\end{figure}

Solving the above complementary conditions we have calculated 
twelve Juddian points for the
resonant TPRH with $2\omega = \omega_0 =1$, 
corresponding to values of of $2 \le N \le 7$.
These are displayed in Table 
\ref{2pJuddtab1}, listed to the first
10 significant figures, and we have used the original units of 
Eq. (\ref{TPRH}). The $N=2$ and $N=3$ cases have such
simple complementary conditions that closed analytic expressions
may be found.  On resonance these are given by
$g=\frac{1}{8\sqrt{2}}$ and
$E=\frac{1}{4}\rb{\frac{5}{\sqrt{2}}-1}$ for $N=2$, and
$g=\frac{1}{8}\sqrt{\frac{3}{10}}$ and
$E=\frac{1}{4}\rb{7\sqrt{\frac{7}{10}}-1}$ for $N=3$.
This set of twelve Juddian points is indicated on the energy
schema of the Hamiltonian in Fig.
\ref{2pJuddfig1}, where the schema was
obtained by approximate numerical diagonalisation via a
standard configuration-interaction method, using a basis
size of the lowest 501 harmonic oscillator states \cite{ce:th}.  The 
energy baselines, $E = \frac{1}{2}
\rb{-\frac{1}{2} + \rb{N + \frac{1}{2}}\Omega}$, are also plotted.

As is clearly seen from Fig. \ref{2pJuddfig1}, the Juddian
solutions found by this method occur at level-crossings in the 
energy schema, but that they do not cover every 
crossing, as is the case with the one-photon Hamiltonian.  
Considering the quantum numbers 
$\pi_{2\gamma}$ of the
two intersecting lines at each crossing, we see that the above
type of solution can only describe the crossings of states having
$\pi_{2\gamma}=+1$ with ones having $\pi_{2\gamma} = -1$, and of
crossings of states having $\pi_{2\gamma} = +i$ with ones having
$\pi_{2\gamma} = -i$.  The remaining four types of possible crossings are
not described.  This situation is summarised in Table
\ref{2pJXings}.

This series of crossings can be understood by considering the
parity operator $\Pi_{2 \gamma}^2$
\begin{equation}
\Pi_{2 \gamma}^2 = \exp \rb{i \pi b^\dagger b},
\end{equation}
which obviously commutes with the Hamiltonian. 
From considering the eigenvalues of this operator we see that the 
Juddian solutions we have
described occur between levels which have the same value of
$\Pi_{2 \gamma}^2$.  Thus although the Ans\"{a}tze of the Juddian
solutions above are not eigenstates of the Fourier-like operator 
$\Pi_{2 \gamma}$, they are eigenstates of parity.

These crossings can be viewed in another way.  Ng {\it et al.} have 
introduced
a unitary transformation which decouples the spin and bosonic degrees 
of freedom \cite{ng:lo}.  After the application of this transformation, the bosonic part
of the Hamiltonian is given by
\begin{equation}
h^{(M)} = \frac{1}{2}M \omega_0 \left(-1\right)^{K_0-k}
  + 2\omega\left(K_0 - \frac{1}{4}\right)
 +4 g \left(K_+ + K_-\right), \label{2p1bodh}
\end{equation}
where the numbers $M=\pm 1$, corresponding to the spin degree of freedom,
and $k = \frac{1}{4}, \frac{3}{4}$, corresponding to the Bargmann index,
serve to characterize the four independent sub-spaces into which
the full Hilbert space of the Hamiltonian decomposes under this 
transformation.  We thus see that the crossings detailed
above occur between states with the same Bargmann $k$ index, but
with different $M$ indices.  By contrast the missing crossings occur
between states with different values of $k$.

The reason why the above Ans\"{a}tze can describe these solutions
and not the others is as follows.  The solutions that we have been
able to find occur at crossings between energy eigenfunctions that
both have the same Bargmann index $k$, which means that both
states are composed of either all even or all odd number states.
At the Juddian points these two eigenstates become degenerate in
energy and thus, to find the energy at the level-crossing we may
form a linear superposition of the two eigenstates, which will,
in general, not be an eigen-state of $\Pi_{2 \gamma}^2$.  
Because the degenerate energy
eigenstates belong to the same $k$-sector, the formation of the
superposition allows the individual terms in one wavefunction to
add to the terms in the other.  If we form the superposition
correctly, the resultant wavefunction may have a form much
simpler than the constituent wavefunctions.  This is exactly the
case when we choose the Ansatz (\ref{2pJSansatz}).

The solutions that we have been unable to find with the above
method occur at the level-crossings between energy eigenstates
that have different Bargmann indices. This means that one
eigenstate is composed of odd number states, whilst the other is
composed only of even number states.  Consequently, no
superposition of these states will lead to a reduction in the
complexity of either wavefunction and we have been unable to find
simple Ans\"{a}tze at these level-crossings.

Although the above method is not directly extensible to the
remaining crossings, it may still be the case that exact
solutions can be found.  Although there is no {\it a priori}
reason to expect that these exact solutions exist, by looking at
the energy schema generated numerically, we observe that the remaining
level-crossings appear to lie along base-lines described by
\begin{equation}
\tilde{E} = -\frac{1}{2} + n \Omega,~~~~~~~n=2,3,\ldots,
\end{equation}
where, as previously, $\Omega = \sqrt{1 - 4 \lambda^2}$.  These
baselines are so similar to the baselines for the Juddian
solutions found above that it would seem likely that Juddian
solutions could also be found at these remaining crossings.

\section{Conclusions}

We have shown that a set of isolated, exact solutions exists
for the two-photon Rabi Hamiltonian.  These are seen to occur at a sub--set 
of the 
level-crossings in the energy schema and although we have not described 
every level crossing, we have been able to explain this in terms of 
the symmetry properties of the crossing states.

This type of isolated solution also occurs in the single-photon Rabi 
Hamiltonian, where they are referred to as Juddian solutions.  Several 
methods have been proposed for finding these solutions \cite{re:nu,ku:le}.  
In {\bf I} we  have given a method for finding these solutions
that is directly analogous 
to the one used here, except that for the one-photon case
we have used displaced, rather than squeezed, bosons.  In this case, the
Ansatz does provide solutions at every level-crossing in the spectrum.  It 
should be noted that the single-photon Hamiltonian  is a simpler 
model than the TPRH, as the 
conserved quantum number analogous to $\pi_{2\gamma}$ only has two 
eigenvalues, and thus there is only one type of level crossing, whereas
in the current model $\Pi_{2\gamma}$ has four eigenvalues and 
there are six different types of crossing.

It is hoped that the results presented here, in conjunction with those in
{\bf I}, will be of use in the analysis of this kind of non-adiabatic model.  
These isolated solutions seem an ideal starting point for the analysis 
of this kind of situation, since they provide exact results which may 
be used as 
bench-marks for further methods.  They also provide crucial insight into 
the symmetries of the model.

\section{Acknowledgments}

One of us (C. E.) acknowledges the financial support of a research studentship
from the Engineering and Physical Sciences Research Council (EPSRC) of
Great Britain.

\begin{table}[p]
\begin{center}
\begin{tabular}{|c|c|c|}
\hline
$N$ & $g$ & $E$ \\
\hline \hline
2 & 0.08838834765 & 0.6338834765 \\
3 & 0.06846531969 & 1.214155046  \\
4 & 0.1136829135  & 0.6855144259 \\
4 & 0.05510006004 & 1.769611501 \\
5 & 0.1017761788  & 1.346571001  \\
5 & 0.04587381623 & 2.308117863  \\
6 & 0.1195668196  & 0.6977617553 \\
6 & 0.09065527261 & 1.987605007  \\
6 & 0.03920841953 & 2.835982030  \\
7 & 0.1124265002  & 1.389132039  \\
7 & 0.03419600455 & 3.356947455  \\
7 & 0.08111783821 & 2.603139795  \\
\hline
\end{tabular}
 \caption{The couplings ($g$), energies ($E$), as well as
order ($N$)of the first twelve Juddian points of the resonant two-photon Rabi
Hamiltonian ($2 \omega = \omega_0 = 1$).\label{2pJuddtab1}}
\end{center}
\end{table}

\begin{table}[t]
\begin{center}
\begin{tabular}{|c||c|c|c|c|}
\hline
~ & $+1$ & $+i$ & $-1$ & $-i$ \\
\hline \hline
$+1$ & - & n & y & n \\
$+i$ & n & - & n & y \\
$-1$ & y & n & - & n \\
$-i$ & n & y & n & - \\
\hline
\end{tabular}
 \caption{Description of the parities of the
level-crossings described by the TPRH Juddian solutions presented
here.  ``y'' denotes that the crossing is described,
``n'' that it is not, whereas ``-'' indicates that no such
crossing occurs.\label{2pJXings}}
\end{center}
\end{table}


\begin{thebibliography}{99}
\bibitem{ii:ra} I. I. Rabi, Phys. Rev. {\bf 51}, 652 (1937).
\bibitem{al:eb} L. Allen and J. H. Eberly, 
  {\it Optical Resonance and Two-Level Atoms}, (Wiley, New York, 1975).
\bibitem{ja:cu} E. T. Jaynes and F. W. Cummings, 
  Proc. IEEE {\bf 51}, 89 (1963).
\bibitem{sh:wa} Y. R. Shen, Phys. Rev. {\bf 155}, 921 (1967); 
   D. F. Walls, J. Phys. A {\bf 4}, 813 (1971).
\bibitem{re:mw}  M. Reid, K. J. McNeil, and D. F. Walls, 
  Phys. Rev. A {\bf 24}, 2029 (1981).
\bibitem{su:bu} C. V. Sukumar and B. Buck, 
  Phys. Lett. A {\bf 83}, 211 (1981); 
  J. Phys. A {\bf 17}, 885 (1984).
\bibitem{co:pe} G. Compagno and F. Persico in 
  ``Coherence and Quantum optics V'',
  edited by L. Mandel and E. Wolf (p.1117, Plenum 1984).
\bibitem{pu:bu} R. R. Puri and R. K. Bullough, 
  J. Opt. Soc. Am. B {\bf 5}, 2021 (1988).
\bibitem{ar:se} A. H. Toor and M. S. Zubairy, 
  Phys. Rev. A {\bf 45}, 4951 (1992)
\bibitem{fa:zh} M. Fang and P. Zhou, 
  J. Mod. Opt. {\bf 42}, 1199 (1995).
\bibitem{cc:ge}  C. C. Gerry, 
  Phys. Rev. A {\bf 37}, 2683 (1988); 
  C. C. Gerry and P. J. Moyer, 
  {\it ibid} {\bf 38}, 5665 (1988).
\bibitem{ge:ba} T. R. Gentile, B. J. Hughey, D. Kleppner, and T. W. Ducas, 
  Phys. Rev. A {\bf 40}, 5103 (1989).
\bibitem{ga:ba} M. Gatze, M. C. Baruch, R. B. Watkins, and T. F. Gallagher, 
  Phys. Rev. A {\bf 48}, 4742 (1993).
\bibitem{ng:lo} K. M. Ng, C. F. Lo, and K. L. Liu, 
  Eur. Phys. J. D {\bf 6}, 119 (1999);  
  C. F. Lo, K. L. Liu, and K. M. Ng, 
  Europhys. Lett. {\bf 42}, 1 (1998).
\bibitem{fe:ra} I. D. Feranchuk, L. I. Komarov, and A. P. Ulyanenkov, 
  J. Phys. A: Math. Gen. {\bf 29}, 4035, (1996).
\bibitem{fo:oc} G. W. Ford and R. F. O'Connell, 
  Physica A {\bf 243}, 377 (1997).
\bibitem{ju:dd} B. R. Judd, 
  J. Chem. Phys. {\bf 67}, 1174 (1977);  
  J. Phys. C: Solid State Phys. {\bf 12}, 1685 (1979).
\bibitem{re:nu} H. G. Reik, H. Nusser, and L. A. Amarante Ribeiro, 
  J. Phys. A: Math. Gen. {\bf 15}, 3491 (1982);  
  M. Ku\'{s}, J. Math. Phys. {\bf 26}, 2792 (1985);  
  H. G. Reik and M. Doucha, Phys. Rev. Lett {\bf 57}, 787 (1986); 
  H. G. Reik, P. Lais, M. E. St\"{u}tzle, and M. Doucha, 
  J. Phys. A: Math. Gen {\bf 20}, 6327 (1987).
\bibitem{gr:ho} R. Graham and M. H\"{o}hnerbach, 
  Phys. Lett. A {\bf 101}, 61 (1984).
\bibitem{ho:ta} G. Hose and H. S. Taylor, 
  Phys. Rev. Lett. {\bf 51}, 947 (1983).
\bibitem{ce:r1} C. Emary and R. F. Bishop, 
  J. Math. Phys. {\bf 43}, 3916 (2002). 
\bibitem{bi:v4}  R. F. Bishop and A. Vourdas, 
  Phys. Rev. A {\bf 50}, 4488 (1994)
\bibitem{pe:re}  A. Perelomov, 
  {\it Generalised Coherent States and Their Applications}, 
  (Springer, Berlin,  1986).
\bibitem{cg:rg} C. C. Gerry and R. Grobe, 
  Phys. Rev. A {\bf 51}, 4123 (1995).
\bibitem{bi:v3} R. F. Bishop and A. Vourdas, 
  Z. Phys. B {\bf 17}, 527 (1988).
\bibitem{bi:v2} R. F. Bishop and A. Vourdas, 
  J. Phys. A: Math. Gen. {\bf 19}, 2525 (1986).
\bibitem{ce:th} C. Emary, PhD Thesis, 
  UMIST, Manchester, 2001 (unpublished).
\bibitem{ma:ph} M. Reed and B. Simon, 
  {\it Methods of Modern Mathematical Physics Vol. 1: Functional Analysis} 
  (Academic, New York 1972).
\bibitem{ku:le}  M. Ku\'{s}, J. Math. Phys. {\bf 26}, 2792 (1985);  
  M. Ku\'{s} and M. Lewenstein, J. Phys. A: Math. Gen. {\bf 19}, 305 (1986).










%  \bibitem{eb:na} J. H. Eberly, N. B. Narozhny and J. J. Sanchez-Mondragon, Phys. Rev. Lett. {\bf 44}, 1323 (1980);  N. B. Narozhny, J. J. Sanchez-Mondragon, and J. H. Eberly, Phys. Rev. A {\bf 23}, 236 (1981);  H-I. Yoo, J. J. Sanchez-Mondragon, and J. H. Eberly, J. Phys. A: Math. Gen. {\bf 14}, 1383 (1981).
% 
%  \bibitem{pu:bu} R. R. Puri and R. K. Bullough, J. Opt. Soc. Am. B {\bf 5}, 2021 (1988).
%  \bibitem{ro:gl} R. Glauber, Phys. Rev. {\bf 131}, 2766 (1963).
%  \bibitem{ba:rg} V. Bargmann, Comm. Pure Applied Math. {\bf 14}, 187 (1964).
%  \bibitem{ng:lo} K. M. Ng, C. F. Lo, and K. L. Liu, Eur. Phys. J. D {\bf 6}, 119 (1999);  C. F. Lo, K. L. Liu, and K. M. Ng, Europhys. Lett. {\bf 42}, 1 (1998). 
%  \bibitem{bi:v4} R. F. Bishop and A. Vourdas, J. Phys. A: Math. Gen. {\bf 20}, 3727 (1987).
%  \bibitem{al:zu} P. Alsing and M. S. Zubairy, J. Opt. Soc. Am. B {\bf 4}, 177 (1987).
%  \bibitem{al:bo}  M. Alexanian and S. K. Bose, Phys. Rev. A {\bf 52}, 2218 (1995).
%  \bibitem{na:ra} T. Nasreen and M. S. K. Razmi, J. Opt. Soc. Am. B {\bf 8}, 2303 (1991).
%  \bibitem{si:ng} S. Singh, Phys. Rev. A {\bf 25}, 3206 (1982).
%  \bibitem{bi:cc} R. F. Bishop, N. J. Davidson, R. M. Quick, and D. M. van der Walt, Phys. Rev. A {\bf 54}, 4657 (1996).
%  \bibitem{bi:va} R. F. Bishop, N. J. Davidson, R. M. Quick, and D. M. van der Walt, Phys. Lett. A {\bf 254}, 215 (1999).
\end{thebibliography}
\end{document}